\documentclass[11pt]{article}
\usepackage[margin=1.15in]{geometry}
\usepackage{amsmath,amssymb}
\usepackage{booktabs}
\usepackage{graphicx}
\usepackage{placeins}
\usepackage[round]{natbib}
\usepackage[colorlinks=true,linkcolor=black,citecolor=black,urlcolor=blue]{hyperref}
\graphicspath{{figures/}}

\newcommand{\degF}{$^\circ$F}
\newcommand{\E}{\mathbb{E}}

\title{Prediction Markets Beat the Weather Forecast on Tomorrow's High Temperature}
\author{Alexander W. Crosier}
\date{September 20, 2026}

\begin{document}
\maketitle

\begin{abstract}
\noindent The sooner we receive information, and the more accurate it is, the 
better planning decisions we can make. Every day, prediction markets let 
anyone bet on tomorrow's high temperature in cities around the world, creating 
a market-implied forecast built on dispersed information. We use the past 
five years of market data from the Kalshi exchange for seven American cities 
to extract, hour by hour, the market-implied forecast. We use this forecast 
as a measuring instrument to see how much information about the temperature 
the market makes public before the public forecasting system does. We race 
it against the leading American and European weather forecasts. In six
of the seven cities we study, the market beats the most accurate single 
public forecast, the National Blend of Models (NBM).
Aggregating every city-day, at the end of the market's first hour of trading it
beats the best single public product by about 10 percent in root-mean-square
error, and holds its lead through the day, overnight, and into the target day. 
Looking at how the forecasts move over time, we find the National Blend
travels four times further toward the market between its postings than the market
travels toward the NBM. The market does not react to new weather forecast updates; 
instead, the forecast slowly publishes information that the market had already shared publicly.
\end{abstract}

\section{Introduction}\label{sec:intro}

Growth in recent years of the exchanges Kalshi and Polymarket has exposed millions 
of people to prediction markets. These platforms have created wide-reaching and 
liquid exchanges that allow us to test
the theorized benefits of prediction markets, including their ability to 
aggregate information and incentivize accurate forecasts. 
Free-market economists for decades have argued that no single individual or institution 
possesses the information needed to make the best economic decisions, and that 
markets can aggregate dispersed information into useful signals. More than 80 years ago, 
Hayek argued that ``knowledge of the circumstances of which we must make use
never exists in concentrated or 
integrated form, but solely as the dispersed bits of incomplete and 
frequently contradictory knowledge which all the separate individuals possess'' \citep{hayek1945}. 
While information aggregation in financial markets has been widely studied, prediction markets 
have historically lacked the scale and liquidity to permit such tests. 
The growth of platforms today now permit such studies, and if prediction markets can aggregate 
dispersed information to create a more accurate forecast than any one institution, their 
implied forecasts should be used for decision making.

Kalshi lists daily markets on tomorrow's high temperature in major 
cities, with markets for some cities starting as far back as 2021. Many industries 
such as airlines, food processing facilities, and electrical utilities use next-day temperature forecasts to 
inform critical and potentially costly operational decisions. For example, an additional 
degree Fahrenheit on a hot summer day can increase demand on a city's power grid by hundreds 
of megawatts; planning to commit that generation a day in 
advance can save tens of thousands of dollars.

In addition to being valuable information, 
next-day temperature forecasts are well suited as a test case for the capabilities of prediction markets. 
Temperature markets on Kalshi consist of a ladder of Yes or No bets such as ``81\degF{} or lower,'' 
``82--83\degF,'' \dots, ``90\degF{} or higher.'' The prices of these contracts, 
when read together, form a probability distribution for tomorrow's
high temperature. The event repeats every day in every city, and we can compare the market-implied 
forecast against a suite of public weather forecasts at each hour those forecasts update. 
The central question we ask is if the market offers some information that the public weather forecasts do not 
themselves contain. Traders can read every public forecast before placing a bet, so
their information set should include what the forecasting systems have published; if the
price is an expectation conditioned on the public information, it should do better 
in the long term than any public forecast. The 
amount by which it does better is the quantity we measure: How much information 
about the temperature have Kalshi traders published that public weather forecasts have not?

\paragraph{Main findings.}
Combining data across cities, the market's root-mean-square error at the end
of its first hour of trading is about 10 percent below the best single
public product, the National Blend of Models (NBM). The market keeps its advantage 
throughout the day as both its forecast and the public guidance become more accurate, and by the
NBM's final posting on
the target morning the Kalshi forecast still has lower error by 
11 percent. 
We construct an optimally weighted combination
of the entire public stack, the forecast of a sophisticated free rider (see Section~\ref{sec:freerider}),
and find that the
market has 3 percent lower error at the opening hour and 4 percent in the
evening on the preceding day.
While the Kalshi forecasts outperform the public forecasts in nearly every city,
the margin is largest in Philadelphia, Chicago, and New York City in winter, when forecasters need 
to time the arrival of cold fronts. The forecast advantage is 
smallest in Miami, where tropical weather conditions lead to lower variation in high temperatures from 
one day to the next. 
Finally, in an event study around the times that the public forecasts are published, we find 
the revisions to the NBM forecast move toward the existing
market forecast about four times as much as the market moves toward the NBM.

\paragraph{Related work.}
Weather has been read out of asset
prices since \citet{roll1984} showed that orange juice futures anticipate
frosts in Central Florida. And temperature derivatives have been shown to outperform 
meteorological forecasts on monthly horizons \citep{ritter2013}. Our study 
mirrors \citet{gurkaynak2007}, who read binary 
options as a probability distribution and race its mean 
against a survey of professional forecasters.
A close design is \citet{cowgill2015}, who race corporate prediction
markets against in-house expert forecasters. In concurrent work, 
\citet{brownP2026wp} compares temperature
markets on the ForecastEx exchange with twenty forecast products at 24 
different locations over seven months of 2026. Both studies find the market 
ahead of every individual product and ahead of a de-biased combination of the weather 
forecasts by a much smaller margin.

\section{The setting}\label{sec:data}

\subsection{What the price measures}\label{sec:measure}

Kalshi offers contracts as a ladder of mutually exclusive intervals for where the 
coming high temperature could fall. A contract pays \$1 if the settled high lands in its interval $I_i$
and \$0 otherwise, so a trader who believes that there is a probability $p$ that the high will land 
in that interval should
buy the contract if its price is below $p$ dollars and sell if it is above. The price $\pi_i$ of contract $i$
reflects the traders' collective probability,
\begin{equation}
\pi_i \;=\; \Pr\!\big(T \in I_i \,\big|\, \mathcal{I}_M\big),
\label{eq:price}
\end{equation}
where $T$ is the day's high at the settlement station and $\mathcal{I}_M$ is
the information held by market participants. We take the mean of that distribution,
\begin{equation}
\mu_M \;=\; \sum_i \pi_i\, c_i
\label{eq:mean}
\end{equation}
as the market's forecast, where $c_i$ is the center of rung $i$.\footnote{The open tails
have no center and so we give them geometrically decaying mass (Appendix~\ref{app:pmf}).
In Appendix~\ref{app:robust} we show all of the findings still hold when reproduced
with the implied median, which does not depend on how the tails are modeled. We also
present liquidity screens in Appendix~\ref{app:screens}.}

\paragraph{A superset of public information.}
Traders can read the public forecasts before placing their bets, so $\mathcal{I}_M$
contains everything the forecasting system has published along with whatever any trader brings, 
whether that is a
proprietary model, further post-processing, or a close read of the Farmers' Almanac. Because the
price is a conditional mean, $\mu_M = \E[T \mid \mathcal{I}_M]$, and a public
forecast $F$ is measurable with respect to $\mathcal{I}_M$, then
\begin{equation}
\E\big[(T-\mu_M)^2\big] \;\le\; \E\big[(T-F)^2\big].
\label{eq:nesting}
\end{equation}
Traders do not need to make a better weather model than the meteorologists. They only need to
start from what has been published and add anything that the weather guidance missed. The gap in
(\ref{eq:nesting}) is therefore how much the public forecasts left on the table at that
moment.

\paragraph{Measuring the gap.}
For a forecast $F$ at a particular time and in a particular city, the error on day $d$ is
$e^{F}_{d} = T_{d} - F_{d}$, and its accuracy over the $n$ days on which
it exists is the root-mean-square error (RMSE),
\begin{equation}
\mathrm{RMSE}_{F} \;=\;
\sqrt{\frac{1}{n}\sum_{d=1}^{n}\big(e^{F}_{d}\big)^{2}}\,.
\label{eq:rmse}
\end{equation}
The market's error is $e^{\text{mkt}}_{d} = T_{d} -
\mu^{\text{mkt}}_{d}$, with $\mu^{\text{mkt}}_{d}$ the implied
mean from~(\ref{eq:mean}) quoted at that hour on day $d$. The market's percentage advantage is therefore
\begin{equation}
\mathrm{gap}_{F} \;=\; 100\,\cdot
\frac{\mathrm{RMSE}_{F} - \mathrm{RMSE}_{\text{mkt}}}{\mathrm{RMSE}_{F}}.
\label{eq:gap}
\end{equation}

\subsection{One contract's life}

\begin{figure}[tp]\centering
\includegraphics[width=\textwidth]{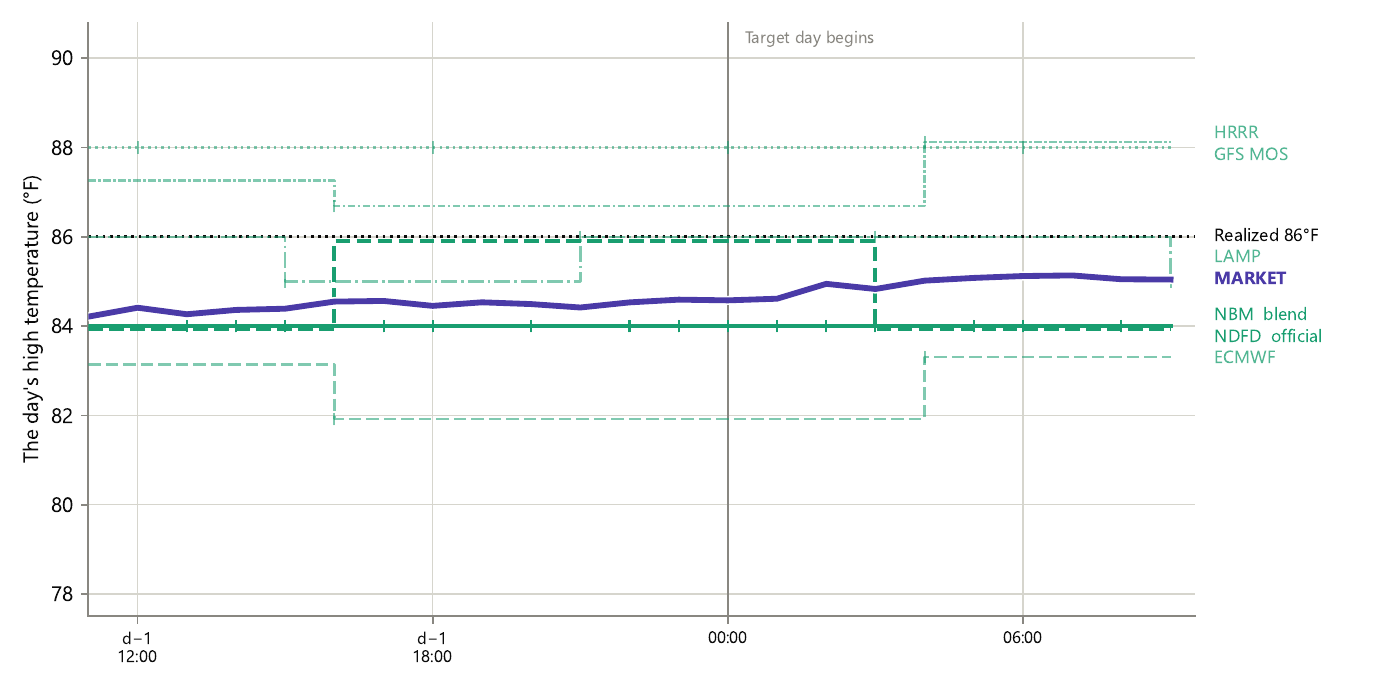}
\caption{One Kalshi contract's life and the published public forecasts for the high 
temperature in New York City on August 1, 2026. The market-implied forecast 
is shown in purple, and the ticks of the green public forecasts show when 
they are published.}\label{fig:life}
\end{figure}

Figure~\ref{fig:life} traces the market-implied mean and the public forecasts 
through an example day, August 1, 2026, in New York City. The Kalshi market 
lists at 10:00 AM Eastern on the day before, and each public forecast updates at 
specific intervals throughout the preceding and target day. The NBM's final 
posting of the day's maximum temperature is published at 8:40 AM Eastern
(12:40 UTC) on the target day, and we end our study at 13:00 UTC.

\subsection{The cities}

We study the seven cities whose markets have existed for more than a year, and they 
experience diverse meteorological conditions. Markets first listed in August 2021 for 
New York and Chicago, May 2023 for Miami and Austin, 
November 2024 for Philadelphia and Denver, and January 2025 for Los Angeles. 
Our city panel includes a total of 7{,}590 city-days, with the last markets taken on
August 12, 2026.\footnote{After this date Kalshi changed its settlement source from the National Weather Service 
to The Weather Company. We choose to keep all contracts used in this study 
settling on the National Weather Service's report.} Each contract
settles on the Weather Service's daily climate report for a named station.

\begin{table}[t]
\caption{The seven cities under study. Each contract settles on the Weather Service's
daily climate report for the city's station.}\label{tab:sample}
\begin{minipage}[c]{0.61\textwidth}\centering
\begin{tabular}{@{}lllr@{}}
\toprule
City & Settlement station & Sample from & Days \\
\midrule
New York & Central Park & Jan 2022 & 1,684 \\
Chicago & Midway & Jan 2022 & 1,684 \\
Miami & Miami Intl & May 2023 & 1,189 \\
Austin & Austin--Bergstrom & May 2023 & 1,186 \\
Philadelphia & Philadelphia Intl & Nov 2024 & 631 \\
Denver & Denver Intl & Nov 2024 & 631 \\
Los Angeles & Los Angeles Intl & Jan 2025 & 585 \\
\midrule
\multicolumn{3}{@{}l}{Total city-days} & 7,590 \\
\bottomrule
\end{tabular}
\end{minipage}\hfill
\begin{minipage}[c]{0.37\textwidth}\centering
\includegraphics[width=\linewidth]{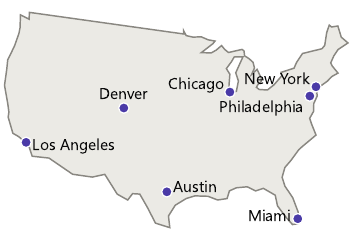}
\end{minipage}
\end{table}

\subsection{The public weather forecasts}\label{sec:ladder}

We use six leading public forecasts, each with 
publication times included in their archives. Five forecasts come from the National Weather Service: 
NDFD, the human-edited official forecast on weather.gov; the
NBM, the calibrated National Blend of Models
\citep{craven2020}; LAMP, the hourly,
observation-conditioned statistical station guidance \citep{ghirardelli2010}; 
GFS MOS, the legacy statistical guidance \citep{glahn1972}; and 
HRRR, the 3-km mesoscale model from the National Oceanic and Atmospheric Administration's 
Global Systems Laboratory. The other forecast is published by the European 
Centre for Medium-Range Weather Forecasts (ECMWF). Raw model output from 
HRRR, ECMWF, and LAMP runs cold at 
certain stations, so we give them a rolling bias correction
estimated only on past data (Appendix~\ref{app:bias}). Three of the six products
forecast the daytime maximum rather than the calendar-day maximum the contract
settles on, which we account for and describe in Appendix~\ref{app:target}. 
The archives and their publication timestamps are
described in Appendix~\ref{app:archives}.

\section{Results}\label{sec:race}

\subsection{Facing the best product at each hour}

\begin{figure}[tp]\centering
\includegraphics[width=.9\textwidth]{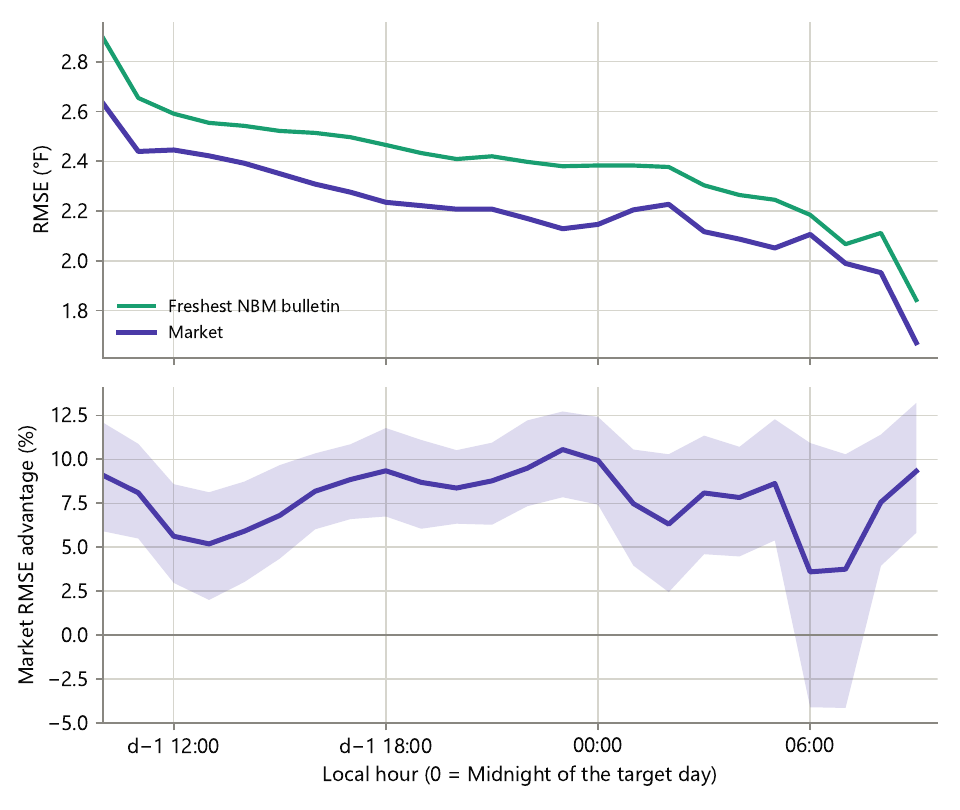}
\caption{Market versus the freshest NBM bulletin at every local hour from
the market's first quoted hour to the NBM's final maximum-temperature
bulletin, seven cities
pooled. Top: RMSE for the NBM (green) and Kalshi (purple). Bottom: The
market's percentage advantage with shading showing 
the 95\% confidence band from bootstrapping.}\label{fig:curve}
\end{figure}

Combining data from each year in every city, we find the market-implied 
forecast has lower error than the best public forecast, the NBM, at 
every hour (Figure~\ref{fig:curve}). At 11:00 AM Eastern, the first 
hour after the market opens, the market's error sits 10 percent below the
freshest NBM's, and the reading holds through the evening, overnight, and into the target morning 
where it sits at 11 percent. As one expects, both forecasts become 
more accurate as the target day approaches: the NBM's error falls from 
2.7\degF{} at the opening
hour to 2.1\degF{} on the target morning, while the market's error falls 
from 2.4\degF{} to 1.9\degF.

\subsection{Three points in time, in depth}\label{sec:freerider}

\begin{figure}[tp]\centering
\includegraphics[width=.95\textwidth]{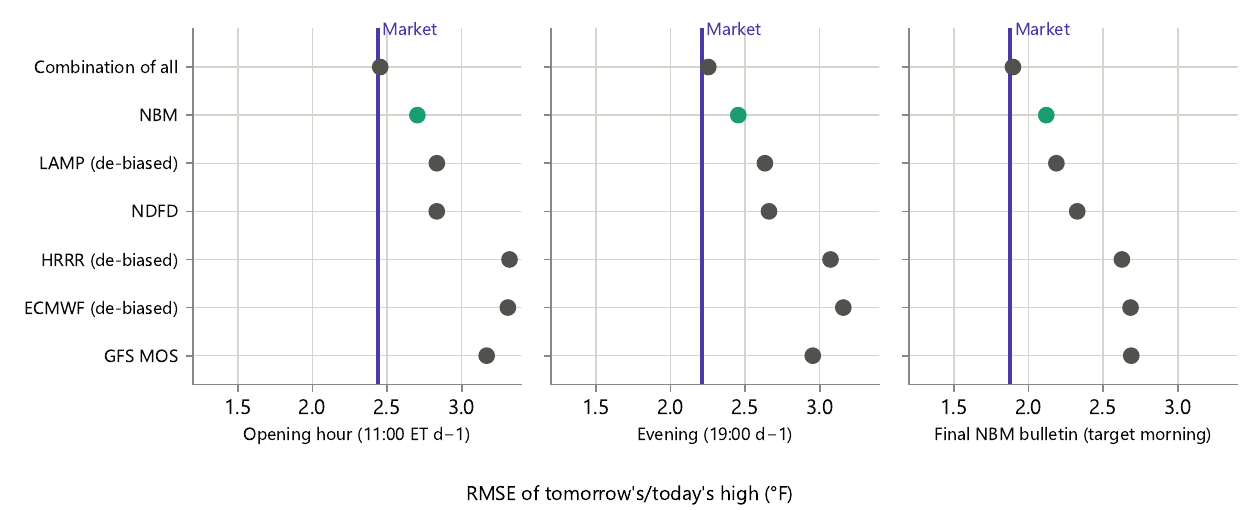}
\caption{RMSE of each public product, their
out-of-sample combination, and the market (vertical purple line) at the three time anchors.}\label{fig:ladder}
\end{figure}

Table~\ref{tab:main} and Figure~\ref{fig:ladder} present errors for the market, each public forecast, and 
an optimal combination of the public forecasts at three points in time: the opening 
hour at 11:00 Eastern the day before (the end of the
market's first hour of trading), the prior evening at 19:00 local time for each contract, 
and the NBM's final bulletin at 13:00 UTC. The market
beats every individual public product at all three anchors. When comparing 
the public guidance, we find the NBM and LAMP are
the strongest products, the official forecast next, and the raw models
last.

\begin{table}[t]\centering
\caption{Market versus the public ladder at the three anchor times. RMSE is 
in $^\circ$F and ``gap \%'' is the market's advantage over
that forecast on the days both exist.
Every gap is significant at 1\% (Appendix~\ref{app:inference}) except the
one marked \textsuperscript{$\ddagger$}, which is significant at 10\% only.}\label{tab:main}
\small\setlength{\tabcolsep}{4pt}
\begin{tabular}{@{}lrrrrrr@{}}
\toprule
 & \multicolumn{2}{c}{Opening hour on d$-$1} & \multicolumn{2}{c}{Evening on d$-$1} & \multicolumn{2}{c}{Final bulletin} \\
\cmidrule(lr){2-3}\cmidrule(lr){4-5}\cmidrule(lr){6-7}
Public benchmark & RMSE & gap \% & RMSE & gap \% & RMSE & gap \% \\
\midrule
Market & \textbf{2.44} & -- & \textbf{2.21} & -- & \textbf{1.88} & -- \\
\midrule
NBM & 2.70 & +9.8 & 2.45 & +10.0 & 2.12 & +11.4 \\
LAMP (de-biased) & 2.83 & +13.9 & 2.63 & +16.3 & 2.19 & +14.2 \\
NDFD & 2.83 & +13.9 & 2.66 & +17.0 & 2.33 & +19.3 \\
GFS MOS & 3.17 & +23.0 & 2.95 & +25.2 & 2.69 & +30.2 \\
HRRR (de-biased) & 3.32 & +26.6 & 3.07 & +28.1 & 2.63 & +28.6 \\
ECMWF (de-biased) & 3.31 & +26.4 & 3.16 & +30.3 & 2.68 & +30.6 \\
\addlinespace
Combination (out-of-sample) & 2.45 & +3.1 & 2.25 & +4.3 & 1.90 & +3.0\textsuperscript{$\ddagger$} \\
\bottomrule
\end{tabular}
\end{table}

In the last row of the table, we report the error of an optimally weighted 
combination of all of the public products,
\begin{equation}
C_d \;=\; \hat{a}_d + \sum_{k}\hat{w}_{k,d}\,F_{k,d},
\qquad
\big(\hat{a}_d,\hat{w}_d\big) \;=\;
\arg\min_{a,\,w}\,\sum_{d'<d}\Big(T_{d'} - a
- {\textstyle\sum_{k}}\, w_k F_{k,d'}\Big)^{2},
\label{eq:combo}
\end{equation}
where $F_{k,d}$ is public product $k$ on day $d$ and the weights are refit 
every day, city by city, on only the days before $d$ (out-of-sample). It exists
on fewer days, where the market's RMSE is 2.38, 2.16, and 1.84. The market
is ahead of this combination by 3.1 percent at the opening
hour and 4.3 percent in the evening. By the final NBM posting 
the gap is 3.0 percent, which our tests find significant only at 10\%. 
The gap across individual public products at these three time anchors ranges from 
9.8\% to 30.6\% as tallied in
Table~\ref{tab:main} and shown in Figure~\ref{fig:ladder}.

Notably, LAMP, which uses live weather observations like temperature and cloud cover 
to update forecasts, drives the public stack's improvement the most over the course 
of the day. If LAMP were dropped from the combination, the market would lead by 3.3, 4.7, and 6.4
percent. In this way the public guidance closes the gap on the market by leaning on real-time 
observations, which may be how market participants beat the public forecasts themselves.

\section{Cities and seasons}\label{sec:mech}

\begin{figure}[tp]\centering
\includegraphics[width=.9\textwidth]{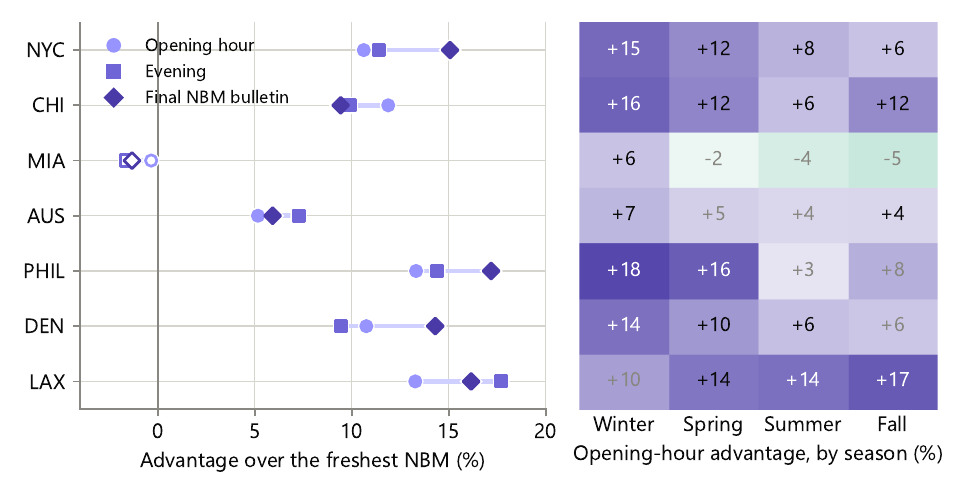}
\caption{Left: The market's advantage over the
freshest NBM by city at the three time anchors. Filled markers are significant at 5\%. Right: the opening-hour
advantage by city and season; grayed values are not significant at 5\%.\label{fig:where}}
\end{figure}

The market has lower error than the NBM in every city and at every time 
anchor, except in Miami. Within cities, the market's advantage is 
higher in some seasons and drops in others, as shown in Figure~\ref{fig:where}.
The market performs better in cities and seasons where the weather forecasts are less accurate. 
In Miami, the NBM's RMSE is 1.6\degF{} on the preceding day's morning, while in 
the other six cities it ranges from 2.6--3.5\degF{}. For most of the year, the Sunshine State
sits under a mass of tropical air, with cold fronts 
rarely reaching South Florida, making forecasting easier compared to the 
mountainous terrain of Denver 
or the frequent cold fronts in Chicago, Philadelphia, and New York. At the opening hour, traders 
stay within 2\degF{} or less of the NBM in Miami on 98 percent of days, while in Chicago 
they stray 2\degF{} or more on 20 percent of days.

We see the same trend when looking into the seasonal behavior. The market's edge is 
largest during the seasons when the forecast error is highest. In November through June when 
the NBM's preceding-day error is 3\degF{} or more, the market beats it by 10 to 16 percent. 
In months when the NBM is already within about two degrees, July through October, the market 
beats it by only 1 to 4 percent. Looking at all cities and months, each additional degree 
of NBM error adds about six percentage points to the market's advantage ($R^2 = 0.22$).

\section{Does the market jump on forecast updates?}\label{sec:event}

In addition to studying accuracy, we can also study how the forecasts move 
over time. The NBM updates its forecast hourly, while the Kalshi markets 
update continuously. When the NBM posts a new forecast, we can ask if the market moves on this 
new information---or if the update actually brings the NBM closer to what the market had already been 
projecting. 
Let $B_t$ be the NBM's
value standing at hour $t$, let $s_t = \mu^{\text{mkt}}_t - B_t$ be the
market's disagreement with it, and, on the hours when an update 
lands, let $R_t = B_{t+1} - B_t$ be the revision to the forecast. The 
following two regressions
ask whether the NBM moves toward the market, and whether the market moves
toward the NBM,
\begin{equation}
R_t \;=\; a + \beta\, s_t + \varepsilon_t,
\qquad\qquad
\mu^{\text{mkt}}_{t+1} - \mu^{\text{mkt}}_t \;=\; a' + \gamma\, R_t
+ \varepsilon'_t.
\label{eq:event}
\end{equation}

We find both coefficients $\beta$ and $\gamma$ are positive and similar in size,
but disagreements are larger than revisions. A typical (one standard deviation)
disagreement of 1.49\degF{} moves the NBM 0.063\degF, while a
typical revision of 0.43\degF{} moves the market 0.016\degF. The NBM travels toward
the market about four times as far as the market travels toward the NBM.

Taking Chicago as an example, on forecast updates where the NBM revised its
forecast by 1\degF{}, the market had moved 0.011\degF{} in
the 45 minutes before the posting and moves merely 0.007\degF{} afterward,
together only 2\% of the forecast's revision, with no jump at the time of publication.
When the new forecast lands,
it does not seem to be news to the traders.

\begin{table}[h]\centering
\caption{Movement from the two regressions:
$\beta_{\text{ant}}$ is NBM's revision per 1\degF{} of market disagreement, while
$\gamma_{\text{react}}$ is the market's move per 1\degF{} of NBM revision. 
$\beta_{\text{gap}}$ is the share of its gap to the market that the
NBM closes by its final bulletin. $t$-statistics are in parentheses.}\label{tab:event}
\begin{tabular}{lrrrr}
\toprule
City & $\beta_{\text{ant}}$ & $\gamma_{\text{react}}$ & $\beta_{\text{gap}}$ & $n$ \\
\midrule
Austin & 0.058 (6.4) & 0.036 (4.4) & 0.225 (4.8) & 17,784 \\
Chicago & 0.035 (8.2) & 0.043 (6.2) & 0.124 (6.4) & 22,241 \\
Denver & 0.052 (6.2) & 0.014 (1.9) & 0.144 (4.3) & 11,205 \\
Los Angeles & 0.042 (15.2) & 0.061 (5.5) & 0.191 (9.2) & 10,362 \\
Miami & 0.079 (18.4) & 0.032 (3.8) & 0.248 (11.9) & 17,371 \\
New York & 0.037 (9.7) & 0.048 (6.6) & 0.129 (8.4) & 22,315 \\
Philadelphia & 0.035 (5.1) & 0.026 (3.6) & 0.155 (4.9) & 11,213 \\
\midrule
Pooled & 0.042 (15.9) & 0.037 (11.9) & 0.153 (13.7) & 112,491 \\
\bottomrule
\end{tabular}
\end{table}

\section{Conclusion}\label{sec:conc}

\begin{figure}[b]\centering
\includegraphics[width=.6\textwidth]{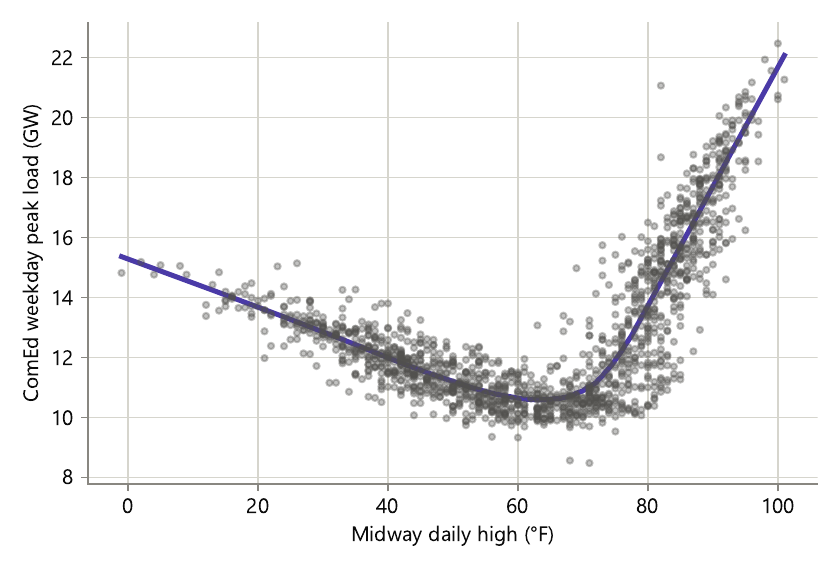}
\caption{ComEd utility peak electrical load against Chicago daily high temperature,
2021--2026.}\label{fig:grid}
\end{figure}

Using temperature markets for seven U.S. cities on the Kalshi exchange, we have shown 
that traders publish a more accurate forecast for tomorrow's high temperature 
than the entire stack of public weather forecasts. At the first hour of trading, the 
Kalshi market has 10 percent lower error than the best single public product. By the 
final posting of the NBM on the target day, the market retains its advantage with 11 
percent lower error. We find the market's advantage is largest in cities and seasons where 
forecasting is hardest, and falls away in Miami where forecast error is lowest. Between
postings, the public forecast drifts toward the market roughly four times
further than the market drifts toward the public forecast, and prices do not 
jump when the National Weather Service publishes.

For any consumer of next-day temperature forecasts, prediction markets offer a public, 
continuously updated, and more accurate forecast than any other public product. 
The strongest example where improved forecasts can have a large economic impact is the 
electricity grid. Taking the Chicago utility zone as an example, each 
additional degree of daily high above 75\degF{} adds roughly 340--480~MW of demand to
the evening peak (Figure~\ref{fig:grid}), so the third of a degree of error
that the market removes from the best public forecast is worth over a hundred megawatts 
of scheduling error, arguably tens of thousands of dollars, on each summer day. Prediction 
markets aggregate knowledge no single forecaster holds, and the information they 
publish should be used for better decision making.

\FloatBarrier

\bibliographystyle{plainnat}
\bibliography{references2}

\clearpage
\appendix
\small

\section{Data and construction}\label{app:data}

\subsection{The market forecast}\label{app:pmf}

Contracts form a ladder of bins
over the high temperature value with an open lower tail, a set of closed intervals, and an
open upper tail. At each snapshot we take the bid--ask midpoint of the market for each
interval in the ladder to get the probability mass for that interval, and spread 
the mass uniformly over the interval. For the open tails, we assign geometrically decaying mass
with ratio $\tau=0.55$ for the next fifteen
integers out, which assigns 95\% of the tail's mass to the nearest five degrees. 
We take the mean of this constructed distribution as the market's forecast.
The mean is sensitive to the tails, so in Appendix~\ref{app:robust} we reproduce every 
headline with the implied median, which does not depend on the tails or $\tau$.

\paragraph{Liquidity screens.}\label{app:screens}
We filter out illiquid markets via three screens: each market must have at 
least three quoted intervals, the bid--ask spread of a market cannot exceed 
50 cents, and the midpoints of quotes, which we take as probabilities, must sum to 
within $\pm$30\% of 1. After screening, we retain 
86\% of observations at the three times under study. Liquidity in the markets 
has improved over time, and almost all of the 
discarded markets were from 2022--2024. Results without the screens
are in Appendix~\ref{app:robust}.

\subsection{Daily highs occurring overnight}\label{app:target}

The Kalshi contracts settle on the maximum temperature recorded for the 
full calendar day. Similarly, for three of the weather products, LAMP, HRRR, and ECMWF, 
we take the maximum over their published hours. 
For the other three products, NBM, NDFD, and GFS MOS, we use their 
explicit daily max temperature field, which technically corresponds to the 
daytime maximum, not including overnight hours. On 4.6\% of days in the 
panel the high falls outside that window. Rather than dropping these days, we 
complete each of these three forecasts with public information about 
the hours it leaves out. That information has 
two parts: LAMP's hourly forecast for the night hours still 
to come, and for comparisons on the target day the highest temperature already observed since local 
midnight. The completed forecast is the highest of the three numbers. 
On an ordinary day, the high temperature is set in the afternoon and nothing changes. 
But on a day when the high was set overnight, this completed forecast 
reflects that, as anyone using the public stack might. 
LAMP, HRRR, and ECMWF already cover the full day, so for 
them we only take the max with the observed temperatures. We also tested the 
results if we drop those days, and we find them unchanged (Appendix~\ref{app:robust}).

\subsection{Public forecast archives}\label{app:archives}

NBM station bulletins and NDFD gridded transmissions come from the NOAA Open
Data Dissemination buckets.
GFS MOS and LAMP come from the Iowa Environmental Mesonet. HRRR and ECMWF come from their
respective open-data archives.

\paragraph{LAMP.}\label{app:lamp}
LAMP is a strong intraday competitor to prediction market forecasts because it 
uses live observations to condition its statistical station guidance. 
Our use of LAMP faces the constraint that the Iowa 
Environmental Mesonet
retains only the 00:00/06:00/12:00/18:00 UTC cycles, not all twenty-four, so the freshest
available LAMP is 2 hours old at the opening hour and 4 hours
in the evening. At the final bulletin the 12:00 UTC LAMP has just landed (median
age twenty minutes).

\paragraph{Bias correction.}\label{app:bias}
HRRR, ECMWF, and LAMP do not publish a daily-high field, so we take each
one's high to be the highest of its hourly 
temperatures. A high built this way runs cold for three reasons. First, 
each hourly value forecasts the expected temperature at that hour,
$\E[T_h]$, and the maximum is a convex function, so by Jensen's inequality
$\max_h \E[T_h] \le \E[\max_h T_h]$: the maximum of the hourly forecasts
is systematically below the expected high. In this case the inequality is strict because it is
uncertain which hour will be the warmest, and the max temperature lands on
whichever hour ends up running warm. Second, the forecasts are valid for 
the hour, while the official high is the peak of a continuous record and
often falls between hours. Third, HRRR and ECMWF forecast an average over
a grid cell, which lacks the granularity for the local conditions at the 
station. LAMP reflects information at the station, 
so it is affected only by the first two. We find that HRRR 
runs cold by up to about 2\degF{}, ECMWF runs cold 1--5\degF{}, and LAMP 
runs cold 1--2\degF{}, depending on the city. We de-bias each using
its historical mean error, estimated by station over the 365 days before the
forecast date. NDFD, the NBM, and MOS publish a maximum-temperature element
of their own, so we use them without correction.

\section{Inference and robustness}\label{app:infrob}

\subsection{Tests}\label{app:inference}

For every city and day we take the market's squared error minus the
benchmark's squared error in our tests of significance. Because some cities 
can experience the same weather systems (like New York, Philadelphia, and Chicago), forecast errors 
are correlated across cities on the same day, so we average
this difference across cities and run every test on the resulting daily
series with one number per date.

To test whether the market is more accurate we use the Diebold--Mariano test
\citep{diebold1995}, a $t$-test of whether the average daily difference is
zero, with Newey--West standard errors to allow for runs of similar days. 
The confidence band in Figure~\ref{fig:curve} comes from a bootstrap that
resamples blocks of ten consecutive dates \citep{kunsch1989}.

\subsection{Robustness}\label{app:robust}

\paragraph{Without the liquidity screens.}
The screens of Appendix~\ref{app:screens} remove 14\% of observations based
on the market's quotes alone, so we check that they do not unduly help our results
by repeating the main comparisons with every observation included. We find the 
weather forecasts do not perform any better on the days we remove than the ones we keep: 
NBM's RMSE is 2.53\degF{} on them and 2.45\degF{} on the rest. Without the
screens, the market's advantage over the NBM is 8.0, 9.0, and 8.0 percent at
the three anchors, compared with 9.8, 10.0, and 11.4 percent with them, and it
remains significant at 1\% at each anchor. The advantage over the combination
of public products is 1.9, 3.2, and 0.7 percent, compared with 3.1, 4.3, and
3.0 percent. It remains significant at the opening hour and in the evening.
At the final bulletin, where the advantage with the liquidity screen is significant only at
10\% (Table~\ref{tab:main}), the advantage without the screen is not significant at
all.

\paragraph{Median instead of mean.}
Every headline is reproduced with the implied median, which is
invariant to how you model the tails (see Appendix~\ref{app:pmf}). The two agree to
within 0.05\degF{} pooled and the gaps at the three anchor times move by less than 
two percent.

\paragraph{Equal weights.}
We also report the equal-weighted average of the same products, since
estimated optimal weights can lose to a simple average in finite samples
\citep{smith2009}. Pooled and on the days both exist, equal
weighting beats the estimated combination at the opening hour (2.43\degF{}
against 2.45), ties it in the evening (2.25 against 2.25), and is a shade
worse at the final bulletin (1.900 against 1.895). Measured against the
equal-weighted average instead, the
market's opening-hour margin is 2.1 percent rather than 3.1 (on these days
the market's RMSE is 2.38\degF{}).

\end{document}